\documentclass[12pt]{article}
\usepackage[english]{babel}
\usepackage[latin1]{inputenc}

\usepackage[intlimits]{amsmath}
\usepackage{amssymb}

\usepackage{url}
\usepackage{graphicx}

\numberwithin{equation}{section}

\DeclareMathOperator{\BR}{BR}
\providecommand{\abs}[1]{\left\lvert #1\right\rvert }
\providecommand{\unit}[1]{\ensuremath{\mathrm{#1}}}
\providecommand{\usk}{\ensuremath{\,}}

\def\lsim{\mathrel{\raise.3ex\hbox{$<$\kern-.75em\lower1ex\hbox{$\sim$}}}}
\def\gsim{\mathrel{\raise.3ex\hbox{$>$\kern-.75em\lower1ex\hbox{$\sim$}}}}
\def\Rp{\not R_p}

\begin{document}

\date{\mbox{ }}

\title{
{\normalsize
DESY 08-122\hfill\mbox{}\\
TUM-HEP 700/08\hfill\mbox{}\\
September 2008\hfill\mbox{}\\}
\vspace{1.5cm} 
\bf Unstable Gravitino Dark Matter\\
and Neutrino Flux \\[8mm]}

\author{Laura Covi$^a$, Michael Grefe$^a$,
Alejandro Ibarra$^b$, David Tran$^{b}$\\[2mm]
{\normalsize\it a Deutsches Elektronen-Synchrotron DESY, Hamburg}\\[-0.05cm]
{\it\normalsize Notkestra\ss{}e 85, 22603 Hamburg, Germany}\\[2mm]
{\normalsize\it b  Physik-Department T30, Technische Universit\"at M\"unchen,}\\[-0.05cm]
{\it\normalsize James-Franck-Stra\ss{}e, 85748 Garching, Germany}
}
\maketitle

\thispagestyle{empty}

\begin{abstract}
\noindent
The gravitino is a promising supersymmetric dark matter candidate
which does not require exact $R$-parity conservation.
In fact, even with some small $R$-parity breaking, gravitinos
are sufficiently long-lived to constitute the dark matter of the Universe,
while yielding a cosmological scenario consistent with primordial
nucleosynthesis and the high reheating temperature required for thermal
leptogenesis. In this paper, we compute the neutrino flux from direct gravitino
decay and gauge boson fragmentation in a simple scenario with
bilinear $R$-parity breaking. Our choice of parameters is motivated 
by a proposed interpretation of anomalies in the extragalactic 
gamma-ray spectrum and the positron fraction in terms of gravitino 
dark matter decay. We find that the generated neutrino flux is compatible 
with present measurements. We also discuss the possibility of detecting 
these neutrinos in present and future experiments and conclude that 
it is a challenging task. However, if detected, this distinctive signal might 
bring significant support to the scenario of gravitinos as decaying dark matter.
\end{abstract}

\newpage

\section{Introduction}

The question of the nature of dark matter is still one of the unsolved mysteries 
in modern cosmology. Many particle candidates have been put forward, but 
until now the only dark matter evidence we have is based on the gravitational 
interaction.
In the context of supersymmetry with conserved $R$-parity, one naturally
encounters one of the most favoured solutions to the dark matter problem.
There, the lightest supersymmetric particle (LSP) is stable and can be
a successful dark matter candidate if it is neutral and weakly interacting like 
the neutralino. The neutralino is the most thoroughly studied dark matter 
candidate and will be tested  in the near future in accelerator, 
direct detection
and indirect detection experiments~\cite{Bertone:2004pz}.

On the other hand, it is also possible that the dark matter interacts only
gravitationally, and supersymmetry also offers candidates of this type.
A prominent example is the gravitino, the superpartner
of the graviton, which was the first supersymmetric dark matter candidate
proposed~\cite{Pagels-Primack}. It is one of the most elusive dark matter 
candidates due to its extremely weak interactions.
In fact, as part of the gravity multiplet, all gravitino interactions
are suppressed either by the Planck scale (for the spin-3/2 component)
or by the supersymmetry-breaking scale (for the Goldstino component).

Usually, having an LSP with such extremely weak interactions
poses a severe problem to Big Bang nucleosynthesis,
as it makes the next-to-lightest supersymmetric particle (NLSP) so long-lived
that it decays during or after the formation of the primordial nuclei,
typically spoiling the successful predictions of the standard
scenario~\cite{BBN-gravitino}. Moreover, if the NLSP is charged,
the formation of a bound state with $^4$He catalyzes the production
of $^6$Li~\cite{Pospelov:2006sc} leading to an overproduction of 
$^6$Li by a factor 300-600~\cite{Hamaguchi:2007mp}.
One way to avoid these constraints is to lower the scale of supersymmetry
breaking, thus enhancing the Goldstino interactions.
However, the reheating temperature of the Universe has to be
lowered accordingly in order to avoid overclosure~\cite{gravitino-TH}
and is then in conflict with the minimal value required by thermal 
leptogenesis in order to explain the baryon asymmetry of the 
Universe~\cite{lepto-TH}.

It was recently proposed in~\cite{bchiy} that these problems are automatically 
solved if a small breaking of $R$-parity is introduced in the model.
Even though in the presence of $R$-parity violation the neutralino LSP
is too short-lived to play the role of dark matter, the gravitino LSP
can still have a sufficiently long lifetime, which is typically
many orders of magnitude greater than the age of the Universe
due to the suppression of the decay rate by the Planck scale
and the small $R$-parity violating couplings~\cite{ty00}.
In this scenario, the NLSP population in the early Universe quickly 
decays into Standard Model particles via $R$-parity violating interactions.
Apart from the presence of a rather inert population of gravitinos, 
produced thermally from superpartner scatterings at reheating, the 
cosmology then reduces to the non-supersymmetric case
well before the synthesis of primordial nuclei.

An intriguing feature of the scenario with $R$-parity breaking is that
gravitino dark matter is not necessarily invisible anymore, since it
will decay into Standard Model particles at a very slow rate.
Since the huge number of gravitinos in our own Galaxy, as well 
as in nearby galaxies and clusters, may compensate for the  highly 
suppressed decay rate, this opens up the possibility of observing the 
dark matter decay products as an anomalous contribution to the diffuse
gamma-ray flux~\cite{bbci07,it07} or the cosmic-ray antimatter fluxes~\cite{it08,imm08}.
Indeed, anomalous excesses have been observed both in the diffuse extragalactic
gamma-ray spectrum and in the positron fraction in similar energy ranges.
It has been pointed out that the decay of gravitino dark matter with a lifetime
of $\sim 10^{26} \usk \unit{s}$ and a mass of
$\sim 150 \usk \unit{GeV}$ can account for both of these excesses
at the same time~\cite{it08,imm08}.
This motivates our study of the corresponding neutrino spectrum for
the same choice of parameters, both as a consistency check and to
find out whether in this scenario an anomalous contribution to the
neutrino flux may be expected in present and future neutrino
experiments.

This paper is organised as follows: in the next section we will briefly
review bilinear $R$-parity violating models and discuss the
resulting gravitino decay modes.
In Section~\ref{Spectrum} we will present the neutrino spectrum
from dark matter decay and describe its main features.
In Section~\ref{Fluxes} we will then give the neutrino flux as a function of the
gravitino lifetime and mass both for neutrinos from our own Galaxy
and from diffuse extragalactic sources and consider the effect of neutrino
oscillations on the signal expected at the Earth.
In Section~\ref{Backgrounds} we will discuss the different neutrino backgrounds
in the energy range we are interested in and compare them to our signal.  
In Section~\ref{Detection} we will propose strategies to disentangle
the signal from the background, compare the result to present neutrino
data from Super-Kamiokande. We will then discuss the feasibility of detection 
in future detectors in Section~\ref{Future} and conclude in Section~\ref{Conclusions}.

\section{Bilinear $R$-parity breaking}
\label{Rparity}

We consider here a model of the type described in detail in~\cite{bchiy},
where the breaking of $R$-parity is related to the breaking of $B-L$.
In this class of models, the lepton number violation in the
superpotential is  encoded in the bilinear term~\cite{hs84}
\begin{equation}\label{bilinear}
W_{\Rp} =  \mu_i L_i H_u \;.
\end{equation}
It can be shown that the $R$-parity violating coupling
$\mu_i$ is suppressed compared to the coefficient $\mu$ of
the $R$-parity conserving bilinear term $H_u H_d$ by
$\mu_i \sim \mu v^2_{B-L}/M_P^2$, with $v_{B-L}$ and $M_P$ being
the scale of $B-L$ breaking and the Planck mass, respectively.
On the other hand, the induced baryon number violation is negligible,
the corresponding Yukawa coupling being further suppressed
by the ratio of the gravitino mass to the Planck mass and a higher power 
of the scale of  $B-L$ breaking 
over the Planck scale, $\lambda^{\prime\prime} \sim m_{3/2} v^4_{B-L}/M_P^5$.
Then, if the scale of $B-L$ breaking is low enough,
the constraints on proton stability are satisfied and
the lepton number violating interactions are kept out of equilibrium
before the electroweak phase transition, thus preventing the erasure
of any previously generated baryon asymmetry.
This condition requires approximately $ \mu_i/\mu \lsim 10^{-6-7} $~\cite{lepto}, which translates into $v_{B-L}\lsim 10^{14}$\usk GeV,
although this could be circumvented for some
specific flavour structures in the Yukawa couplings.
Since the present bounds on the gravitino lifetime are already constraining
the parameters at this order from observations of both gamma rays~\cite{bbci07, it07} and antiprotons~\cite{it08,imm08},
this last option is probably not viable.

Apart from the supersymmetric term above, the corresponding soft bilinear
super-symmetry-breaking term $B_i \mu_i {\widetilde L_i} H_u $ arises in the Lagrangian.
Since the $B_i$ and $\mu_i $ terms are not usually aligned at the
weak scale, a non-vanishing vacuum expectation value ({\it v.e.v.})
is generated along the sneutrino field direction explicitly breaking 
lepton number and generating not only one neutrino mass, but also 
non-vanishing mixings between neutralinos and neutrinos, as well 
as between charginos and charged leptons~\cite{Rp-bilinear}.
Such mixings are responsible for the two-body decays of the gravitino
into gauge boson and neutrino, which are the main source of neutrino
flux in our scenario. These decays are also possible at the one-loop 
level if only trilinear $R$-parity breaking terms are considered, and even in this case 
they can dominate in part of the parameter space~\cite{lor07}.
Since the neutralino--neutrino mixing takes place along the Zino 
component, the branching ratios into the different gauge boson 
channels are fixed by the neutralino mixing matrix
once the gravitino mass is specified.

If the gravitino is lighter than the massive gauge bosons, the dominant decay
channel is the two-body decay into monoenergetic photon and neutrino.
Since all the observed Yukawa couplings are largest for the third
generation, it is reasonable to assume that the $R$-parity breaking 
couplings are also largest for the third generation. Thus, the
sneutrino acquires a {\it v.e.v.} only along the $\widetilde\nu_\tau $ 
direction, and the gravitino will dominantly decay into neutrinos
with $\tau$ flavour.
We will see later that this is not a crucial assumption, since neutrino
oscillations change any pure neutrino flavour into a mixed state.
In particular, due to maximal atmospheric mixing, the flux of
tau and muon neutrinos turns out to be identical.

If instead the gravitino is heavier than the
electroweak gauge bosons, the decay into these particles is
favoured, since the sneutrino has electroweak charge.
In this case, neutrinos are produced not only
in the direct decay into $ Z^0 \nu_{\tau} $, but also in the fragmentation
of the $W^{\pm}$ and $Z^0$ bosons and the decay of the 
$\tau^{\mp}$ leptons, thus adding a continuous component
to the spectrum. In this paper, we will mostly consider the second case
of a heavier gravitino, motivated by the interpretation of anomalies
in other channels as explained above. Furthermore, at lower gravitino masses,
the detection of any signal is much more difficult due to the lower
neutrino yield and the smaller signal-to-background ratio.

\subsection{Gravitino Decay}

In the models with bilinear $R$-parity breaking and a non-zero sneutrino
{\it v.e.v.} along the $\widetilde\nu_\tau $ direction, the main
decay channels for the gravitino are:
\begin{equation}
\begin{split}
\psi_{3/2} &\rightarrow\gamma \nu_{\tau}\,, \\
\psi_{3/2} &\rightarrow W^{\pm} \tau^{\mp}\,, \\
\psi_{3/2} &\rightarrow Z^0 \nu_{\tau}\,, \\
\psi_{3/2} &\rightarrow h \nu_{\tau}\,.
\end{split}
\end{equation}
The first decay is practically always allowed, while the next two
are open only for a gravitino mass above the threshold for 
$W^\pm$ or $Z^0$ production.
At even higher masses, the decay into Higgs boson and neutrino opens up
via the Higgsino component of the neutralino and the $R$-parity violating
Higgs--sneutrino mixing.
We will consider here the case of a large Higgs mass parameter $\mu $,
where the lightest Higgs is Standard Model-like and the other Higgses decouple.

The decay widths for these processes can be computed from the interaction
Lagrangian of a gravitino with a gauge boson and a gaugino or with the two
chiral Higgs multiplets with the insertion of a sneutrino {\it v.e.v.}
~\cite{SUGRA-FR}.
The decays into electroweak gauge bosons arise both from 3--vertices 
and from the
non-abelian 4--vertex, so their structure is more complicated than for
the abelian sector. The results depend on the gaugino mixing
matrices and are given by the following:
\begin{equation}
\begin{split}
\Gamma(\psi_{3/2}\rightarrow\gamma \nu_{\tau})
&=\frac{\xi_\tau^2 m_{3/2}^3}{64 \pi M_P^2}
\abs{U_{\widetilde{\gamma} \widetilde Z}}^2, \\
\Gamma(\psi_{3/2}\rightarrow Z^0 \nu_{\tau})
&=\frac{\xi_\tau^2 m_{3/2}^3}{64 \pi M_P^2}\,\beta_Z^2
\left[  \abs{U_{\widetilde{Z} \widetilde Z}}^2 f_Z
- \frac{8}{3}\frac{ M_Z}{ m_{3/2}}\, \text{Re}\!\left[ U_{\widetilde{Z} \widetilde Z }\right] j_Z
+ \frac{1}{6}\, h_Z \right], \\
\Gamma(\psi_{3/2}\rightarrow W^{\pm} \tau^{\mp})
&=\frac{\xi_\tau^2 m_{3/2}^3}{32 \pi M_P^2}\,\beta_W^2
\left[  \abs{U_{\widetilde{W} \widetilde{W}}}^2 f_W
- \frac{8}{3} \frac{ M_W}{ m_{3/2}}\, \text{Re}\!\left[ U_{\widetilde{W} \widetilde W}\right] j_W
+ \frac{1}{6}\, h_W \right],
\end{split}
\label{decay-widths}
\end{equation}
with $ \xi_\tau = \langle \widetilde\nu_\tau \rangle/v $, where
$ v = 174\usk\unit{GeV} $ is the Higgs {\it v.e.v.}. Assuming $\xi_\tau \ll 1$,
the mixing matrix elements are given by
\begin{equation}
 \begin{split}
U_{\widetilde{\gamma} \widetilde Z}
&\simeq M_Z \sum_{\alpha = 1}^4 \frac{c_{\widetilde\gamma \chi_\alpha} c^{*}_{\widetilde Z \chi_\alpha}}{M_\alpha} \; ,\\
U_{\widetilde{Z} \widetilde Z}
&\simeq M_Z \sum_{\alpha = 1}^4 \frac{c_{\widetilde Z \chi_\alpha} c^{*}_{\widetilde Z \chi_\alpha}}{M_\alpha} \; ,\\
U_{\widetilde{W} \widetilde W}
&\simeq \frac{M_W}{2} \sum_{\alpha = 1}^2
\frac{c_{\widetilde W^{+} \chi_\alpha^{+}} c^{*}_{\widetilde W^{-} \chi_\alpha^{-}}+ h. c.}{M_\alpha^{\pm}} \; .
 \end{split}
\end{equation}
Here, $ c_{ij} $ are the elements of the unitary matrices that diagonalise
the neutralino/chargino mass matrices.
The kinematical factors are given by
\begin{equation}
 \begin{split}
\beta_i &= 1 - \frac{M_i^2}{m_{3/2}^2}\;, \\
f_i &= 1+\frac{2}{3} \frac{M_i^2}{m_{3/2}^2} +\frac{1}{3} \frac{M_i^4}{m_{3/2}^4}\;, \\
j_i &= 1+\frac{1}{2} \frac{M_i^2}{m_{3/2}^2} \;,\\
h_i  &= 1+ 10 \frac{M_i^2}{m_{3/2}^2} + \frac{M_i^4}{m_{3/2}^4}\;.
 \end{split}
\end{equation}
The non-abelian 4--vertex computed here was previously neglected in~\cite{it07}.
Note that our results for the decay rates do not agree exactly with
those given in~\cite{imm08}: for the interference terms
proportional to $j_i$ we find a negative sign and a larger
coefficient. As a result, the branching ratios into
the massive gauge boson channels are slightly smaller
than in~\cite{imm08}, while the branching ratio into photon--neutrino
is larger.

The decay into the lightest Higgs, on the other hand, is given by
\begin{equation}
\Gamma(\psi_{3/2}\rightarrow h \nu_{\tau}) =\frac{\xi_\tau^2 m_{3/2}^3}{ 384 \pi M_P^2}\,\beta_h^4
\abs{ U_{\widetilde{H}_u \widetilde Z}  \sin\beta +
U_{\widetilde{H}_d \widetilde Z}  \cos\beta
+ \frac{m_{\widetilde\nu_\tau}^2}{m_{\widetilde\nu_\tau}^2 - m_h^2} }^2,
\label{width-higgs}
\end{equation}
in the limit where the lightest Higgs is given by $ h =
\sqrt{2} \left( \text{Re} [H_u] \sin\beta + \text{Re} [H_d] \cos \beta \right) $, with
\begin{equation}
U_{\widetilde{H}_i \widetilde Z}
\simeq M_Z \sum_{\alpha = 1}^4 \frac{c_{\widetilde H_i \chi_\alpha} c^{*}_{\widetilde Z \chi_\alpha}}{M_\alpha} \; ,
\end{equation}
and expressing the soft supersymmetry-breaking Higgs--sneutrino mixed mass term
through the sneutrino {\it v.e.v.} as $ m_{\widetilde\nu_\tau}^2 \xi_\tau/\sqrt{2} $.

Note that since the decay into the Higgs is strongly suppressed by the phase
space
factor $\beta_h$, and since we will consider in the following mostly a
gravitino mass of  $150 $\usk GeV, while $ m_h > 114 $\usk GeV for a Standard
Model-like Higgs,
this channel is negligible in our case, as can be seen from Fig.~\ref{BR-fig}.
We will therefore ignore it in the following discussion, although it will be
included in the numerical results and in the figures.

\section{Neutrino Energy Spectrum}
\label{Spectrum}

The injection spectrum of neutrinos from gravitino decay
is composed of a series of contributions. Firstly,
the two-body gravitino decay into a photon and a tau neutrino
produces a monoenergetic line at half the gravitino mass:
\begin{equation}
\frac{dN_{\nu_{\tau}}}{dE}\left( \psi_{3/2}\rightarrow\gamma \nu_{\tau}\right) \simeq\delta\left( E-\frac{m_{3/2}}{2}\right) .
\end{equation}
Additionally, the decay into $Z^0 \nu_{\tau}$
produces a second line at an energy
\begin{equation}
E_{\nu_{\tau}Z}=\frac{m_{3/2}}{2}\left(1 -\frac{M_Z^2}{m_{3/2}^2}\right) ,
\end{equation}
which is not completely monoenergetic due to the
width of the $Z^0$ boson. Instead, it is described by a
normalised Breit--Wigner profile:
\begin{equation}
\frac{dN_{\nu_{\tau}Z}}{dE}
=\frac{1}{\left( E^2-E_{\nu_{\tau}Z}^2\right) ^2+E_{\nu_{\tau}Z}^2\Gamma_{\nu_{\tau}Z}^2}\left( \int_{0}^{\infty}\frac{dE}{\left( E^2-E_{\nu_{\tau}Z}^2\right) ^2+E_{\nu_{\tau}Z}^2\Gamma_{\nu_{\tau}Z}^2}\right) ^{-1},
\end{equation}
where
\begin{equation}
\Gamma_{\nu_{\tau}Z}=\abs{\frac{\partial E_{\nu_{\tau}Z}}{\partial M_Z}}\Gamma_Z=\frac{M_Z}{m_{3/2}}\,\Gamma_Z\,.
\end{equation}

\begin{figure}
\centering
\includegraphics[scale=1.2]{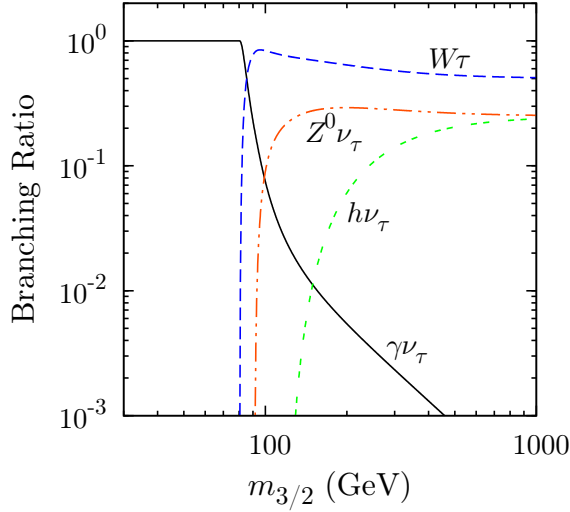}
\caption{Branching ratios of the different gravitino decay
channels as a function of the
gravitino mass for the case of large $\mu $ parameter
(here taken as $\mu = 10 $\usk TeV) and decoupling of the heavy Higgses.
The light Higgs mass is taken to be 115\usk GeV.
The gaugino mass parameters are taken so as to satisfy the unification
relation and to give  $M_1 = 1.5\, m_{3/2} $ at the electroweak scale.
The sneutrino mass parameter is fixed at $ 2\, m_{3/2} $ and
$\tan\beta = 10 $. We see that above the $Z^0$ mass threshold, the
dominant line from $Z^0 \nu_\tau $ always has a branching ratio larger
than 25\usk\%.}
\label{BR-fig}
\end{figure}
Furthermore, the decay into $h \nu_{\tau}$ produces a similar
line, with a differential energy spectrum denoted by $dN_{\nu_{\tau}h}/dE$.
This line is less prominent due to the suppressed branching
ratio into the Higgs decay channel.
Lastly, the fragmentation of the massive gauge bosons produces
a continuous spectrum of neutrinos in all flavours.
We have simulated the fragmentation of the gauge
bosons with the event generator PYTHIA 6.4~\cite{Sjostrand:2006za}
and extracted the spectra in the different neutrino flavours for the
$ W^{\pm} $, $ Z^0 $ and $ h $ channels, which we denote by
$ dN_{\nu_x}^W/dE $, $ dN_{\nu_x}^Z/dE $ and $ dN_{\nu_x}^h/dE $, respectively.
The leptonic decays  $W \rightarrow l \nu$, $Z^0\rightarrow \nu \nu$
and $h \rightarrow l \nu$ also produce,
in the rest frame of the decaying particle,
monoenergetic neutrinos in all flavours. However, due to the boost of the
gauge bosons in different directions, the lines smear out
almost completely in the Earth's rest frame, giving just an
additional contribution to the continuous part of the spectrum.
Taking the various decay channels into account, the total spectra
for the different neutrino flavours are given by:
\begin{equation}
\begin{split}
\frac{dN_{\nu_e}}{dE} =
&\:\BR(W^{\pm} \tau^{\mp})\,\frac{dN_{\nu_e}^W}{dE}+
\BR(Z^0 \nu_{\tau})\,\frac{dN_{\nu_e}^Z}{dE} +
\BR(h \nu_{\tau})\,\frac{dN_{\nu_e}^h}{dE}\,,\\
\frac{dN_{\nu_{\mu}}}{dE} =
&\:\BR(W^{\pm} \tau^{\mp})\,\frac{dN_{\nu_{\mu}}^W}{dE}+
\BR(Z^0 \nu_{\tau})\,\frac{dN_{\nu_{\mu}}^Z}{dE} +
\BR(h \nu_{\tau})\,\frac{dN_{\nu_{\mu}}^h}{dE}\,, \\
\frac{dN_{\nu_{\tau}}}{dE} =
&\:\BR(\gamma \nu_{\tau})\,\delta\left( E-\frac{m_{3/2}}{2}\right)
+ \BR(Z^0 \nu_{\tau})\,\frac{dN_{\nu_{\tau}Z}}{dE}
+ \BR(h \nu_{\tau})\,\frac{dN_{\nu_{\tau}h}}{dE}\\
&+\BR(W^{\pm} \tau^{\mp})\,\frac{dN_{\nu_{\tau}}^W}{dE}
+\BR(Z^0 \nu_{\tau})\,\frac{dN_{\nu_{\tau}}^Z}{dE}
+\BR(h \nu_{\tau})\,\frac{dN_{\nu_{\tau}}^h}{dE}\,.
\end{split}
\end{equation}

\begin{table}
  \centering
  \begin{tabular}{rcccc}
   \hline
   $ m_{3/2}\quad $ & $ \BR(\gamma \nu_{\tau}) $ & $
\BR(W^{\pm} \tau^{\mp}) $ & $ \BR(Z^0 \nu_{\tau}) $ & $
\BR(h \nu_{\tau}) $ \\
   \hline
  $ 10\usk\unit{GeV} $ & 1 & --- & --- & --- \\
  $ 85\usk\unit{GeV} $ & 0.53 & 0.47 & --- & --- \\
  $ 100\usk\unit{GeV} $ & 0.08 & 0.83 & 0.09 & --- \\
  $ 150\usk\unit{GeV} $ & 0.01 & 0.70 & 0.28 & 0.01 \\
  $ 250\usk\unit{GeV} $ & 0.003 & 0.60 & 0.29 & 0.11 \\
  $ 1000\usk\unit{GeV} $ & 0.0002 & 0.51 & 0.25 & 0.24 \\
   \hline
  \end{tabular}
  \caption{Branching ratios into the different gravitino decay channels
for a number of specific gravitino masses. 
The various parameters are chosen as in Fig.~\ref{BR-fig}.}
  \label{BRtable}
\end{table}
The branching ratios for the different decay channels can be
straightforwardly computed from the decay widths, Eq.~(\ref{decay-widths}) and 
(\ref{width-higgs}).
They turn out to depend mainly on the gravitino mass,
with a milder dependence on the ratio between the gaugino masses
$M_1$ and $M_2$ at the electroweak scale.
For illustration, we show in Fig.~\ref{BR-fig} the branching
ratios as a function of the gravitino mass
for the case of a large $ \mu $ parameter, unified gaugino masses
with $ M_1 = 1.5\, m_{3/2} $ and $ m_{\widetilde\nu} = 2\, m_{3/2} $ at the
electroweak scale, and $\tan\beta =10$.
In addition, we list in Tab. \ref{BRtable}, the branching ratios
into the different decay channels for a number of specific gravitino masses.
Note also that in all of the parameter space above the $Z^0$ threshold, at least
two neutrino lines are present with more than 1\usk\% branching ratio.
Actually, sufficiently above the Higgs threshold, the neutrino lines 
from the Higgs and $Z^0$ decay channels have comparable strength. 

\begin{figure}
\centering
\includegraphics[scale=1.2]{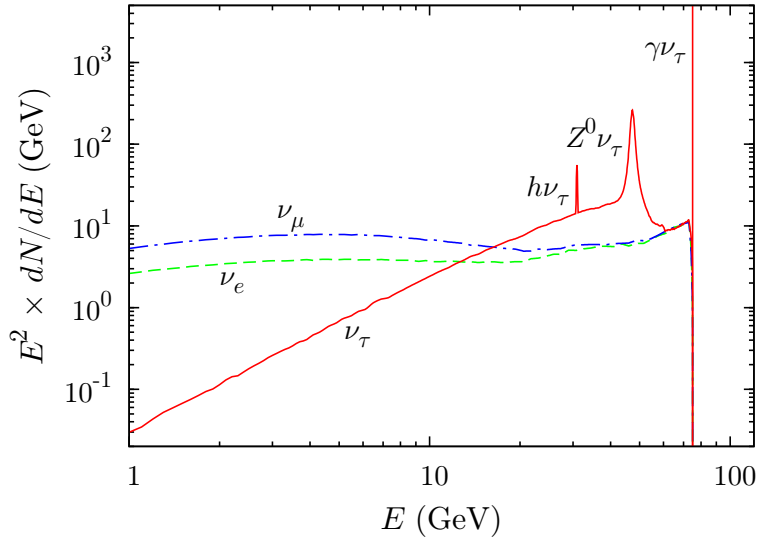}
\caption{Neutrino injection energy spectra from gravitino decay for the
different flavours in the case of $ m_{3/2}=150\usk\unit{GeV} $.
The $Z^0$ line is clearly visible at an energy of 47\usk GeV.
We consider here a Higgs mass of 115\usk GeV, giving a peak 
at $\sim 30 $\usk GeV.}
\label{energyspec}
\end{figure}
The energy spectra for the different neutrino flavours, which were obtained as
described above, are shown in Fig.~\ref{energyspec}.
In this case, a gravitino mass of $ m_{3/2}=150 $\usk GeV was used,
and for that value, all three lines mentioned above are visible in
the spectrum.
We see that the tau neutrino spectrum shows a very characteristic signature
of two or three distinctive peaks (three in the region above the
lightest Higgs threshold, as in this case) in addition to a continuum which is
suppressed at low energies.
The spectra of the other two flavours are very similar to each other,
consisting only of a continuum contribution practically
following a power law behaviour $\propto E^{-2}$ below
the sharp threshold at half the gravitino mass.

\section{Neutrino Fluxes}
\label{Fluxes}

In this section we will consider the diffuse neutrino background
at the Earth arising from the neutrino spectrum from
gravitino dark matter decay as discussed above. The diffuse neutrino flux
has two sources: the decay of gravitinos at cosmological distances and the
decay of gravitinos in the Milky Way halo. The former contribution is
perfectly isotropic, while the latter has a mild dependence on
the Galactic coordinates~\cite{bbci07, it07}.

The decay of gravitinos at a comoving distance $\chi(z)$, where $z$
denotes redshift, produces a neutrino flux with a redshifted energy spectrum
$dN_\nu/d(yE)$, with $y=1+z$. Making use of the variation
of comoving distance with respect to redshift in a matter-
and dark energy-dominated Universe,
$d\chi/dz=(1+z)^{-3/2}/(a_0H_0\sqrt{\Omega_M(1+\kappa(1+z)^{-3})})$,
with  $ a_0 $ and $ H_0 $ being the present cosmic scale factor and the Hubble
parameter, respectively, and $ \kappa=\Omega_{\Lambda}/\Omega_M\simeq3$ being
the ratio between the vacuum and matter density parameters,
it is straightforward to show that the neutrino flux received
at the Earth reads:
\begin{equation}
\frac{dJ_{\text{eg}}}{dE} \simeq A_{\text{eg}}\int_1^{\infty}dy\frac{dN_{\nu}}{d(yE)}\frac{y^{-3/2}}{\sqrt{1+\kappa y^{-3}}}\;,
\label{eg-flux}
\end{equation}
where most neutrinos come from very low redshifts. In this equation,
\begin{equation}
A_{\text{eg}} = \frac{\Omega_{3/2}\rho_c}{4\pi\tau_{3/2}m_{3/2}H_0\Omega_M^{1/2}}
=1.1\times 10^{-7}\usk(\unit{cm^2\usk s\usk sr})^{-1}
\left( \frac{\tau_{3/2}}{1.3\times 10^{26}\usk\unit{s}}\right) ^{-1}
\left( \frac{m_{3/2}}{150\usk\unit{GeV}}\right)^{-1},
\end{equation}
where we have taken the gravitino density to be equal to the cold dark
matter density, $ \Omega_{3/2}h^2=0.1 $, and the other constants
are the critical density $ \rho_c=1.05\, h^2\times 10^{-5}\usk\unit{GeV\usk cm^{-3}} $,
the matter density parameter $ \Omega_M=0.25 $ and $ H_0=100\, h\usk\unit{km\usk s^{-1}\usk Mpc^{-1}} $
with $ h=0.73 $.

In addition to the extragalactic signal there exists a
slightly anisotropic neutrino flux stemming from the decay
of gravitinos in the Milky Way halo. The energy spectrum is given by
\begin{equation}
\frac{dJ_{\text{halo}}}{dE}=A_{\text{halo}}\,\frac{dN_{\nu}}{dE},
\end{equation}
where the intensity of the flux, $ A_{\text{halo}} $, depends on the direction
of observation. It is proportional to the line-of-sight integration over
the halo density profile $\rho_{\text{halo}} $, being defined as
\begin{equation}
A_{\text{halo}}=\frac{1}{4\pi\tau_{3/2}m_{3/2}}
\int_{\text{l.o.s.}}\rho_{\text{halo}}(\vec{l})d\vec{l}\; .
\end{equation}
For our numerical analysis, we will adopt the spherically symmetric
Navarro, Frenk and White (NFW) profile~\cite{Navarro:1995iw}:
\begin{equation}
\rho_{\text{halo}}(r)=\frac{\rho_0}{(r/r_c)
[1+(r/r_c)]^2}\;,
\end{equation}
with $\rho_0=0.26\usk\unit{GeV}/\unit{cm}^3$ and $r_c=20\usk\unit{kpc}$.
The normalisation is chosen such that $\rho(r_\odot) = 0.3$\usk GeV/cm$^3$,
where $r_\odot = 8.5$\usk kpc is the distance of the Sun from the Galactic
centre.
Our conclusions will turn out to be rather insensitive to the
particular choice of the halo profile due to the linear dependence of
the neutrino fluxes on the dark matter density along the line-of-sight
and the fact that we
will integrate the signal over the whole sky excluding the Galactic disk.

After being produced in gravitino decays, neutrinos propagate
while undergoing flavour oscillations. Since neutrinos typically travel
very long distances before reaching us, the conversion probabilities
are~\cite{Strumia:2006db}:
\begin{equation}
\begin{split}
P(\nu_e\leftrightarrow\nu_{\mu}) &=\frac{1}{2}\,
(s_{23}^2\sin^22\theta_{13}+c_{23}^2\sin^22\theta_{12})\,, \\
P(\nu_e\leftrightarrow\nu_{\tau}) &=\frac{1}{2}\,
(c_{23}^2\sin^22\theta_{13}+s_{23}^2\sin^22\theta_{12})\,, \\
P(\nu_{\mu}\leftrightarrow\nu_{\tau}) &=\frac{1}{2}\,
(c_{13}^4\sin^22\theta_{23}-s_{23}^2c_{23}^2\sin^22\theta_{12})\,,
\end{split}
\end{equation}
while the survival probabilities are
\begin{equation}
\begin{split}
P(\nu_e\leftrightarrow\nu_e) &=1-\frac{1}{2}\,
(\sin^22\theta_{13}+c_{13}^4\sin^22\theta_{12})\,, \\
P(\nu_{\mu}\leftrightarrow\nu_{\mu}) &=1-\frac{1}{2}\,
(4\,c_{13}^2s_{23}^2(1-c_{13}^2s_{23}^2)+c_{23}^4\sin^22\theta_{12})\,, \\
P(\nu_{\tau}\leftrightarrow\nu_{\tau}) &=1-\frac{1}{2}\,
(4\,c_{13}^2c_{23}^2(1-c_{13}^2c_{23}^2)+s_{23}^4\sin^22\theta_{12}) \,,
\end{split}
\end{equation}
with $s_{ij} \equiv \sin\theta_{ij}$ and $c_{ij} \equiv \cos\theta_{ij}$. Inserting
into these equations the experimental best fit values for the neutrino
mixing angles $\sin^2\theta_{12}=0.304$, $\sin^2\theta_{23}=0.5$
and $\sin^2\theta_{13}=0.01$~\cite{Schwetz:2008er}, we finally obtain
\begin{equation}
\begin{split}
P(\nu_e\leftrightarrow\nu_e) &=0.56\,, \\
P(\nu_e\leftrightarrow\nu_{\mu})=P(\nu_e\leftrightarrow\nu_{\tau}) &=0.22\,, \\
P(\nu_{\mu}\leftrightarrow\nu_{\mu})=P(\nu_{\mu}\leftrightarrow\nu_{\tau})=P(\nu_{\tau}\leftrightarrow\nu_{\tau}) &=0.39\,.
\end{split}
\label{oscprob}
\end{equation}
Thus, even when the primary neutrino flux is originally mainly composed
of tau neutrinos, the flavour oscillations during the
propagation will produce comparable fluxes of electron, muon
and tau neutrinos due to the large neutrino mixing angles. In particular,
due to the maximal atmospheric mixing angle, the fluxes of muon and
tau neutrinos are expected to be essentially identical.

\begin{figure}
\centering
\includegraphics[scale=1.2]{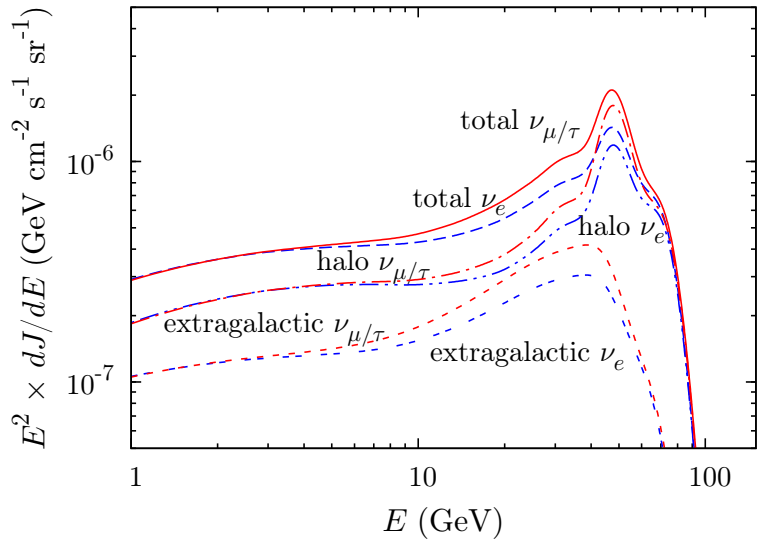}
\caption{Total neutrino fluxes, as well as the extragalactic and halo
contributions
for the different neutrino flavours after propagation using an
energy resolution of $ 10\usk\% $.
The gravitino mass and lifetime are chosen to be $m_{3/2}=150\usk\unit{GeV}$ and $\tau_{3/2}\simeq 10^{26}\usk\unit{s}$.}
\label{eghalo}
\end{figure}

The fluxes for the different neutrino flavours and their extragalactic
and halo contributions are shown in Fig.~\ref{eghalo}. In this
plot, a band of $\pm 10^\circ$ around the Galactic disk has been removed,
and the spectrum is shown with a 10\usk\% energy resolution in order to take the
finite energy resolution of the detector into account.
Note that even with this optimistic assumption for the energy resolution,
the lines from the decay into $\gamma \nu_\tau$ and $h \nu_\tau$ become practically
indistinguishable from the
continuum, whereas the line from the decay into $Z^0 \nu_\tau$ can be resolved.
Its position could allow a determination of the gravitino mass, even
without determining the endpoint of the spectrum.

\section{Neutrino Backgrounds}
\label{Backgrounds}

The detection of a possible neutrino signal from gravitino decay
is hindered by considerable neutrino backgrounds. Namely,
in the energy range of interest, there exist large background neutrino
fluxes produced by interactions of cosmic rays with the
Earth's atmosphere or with the solar corona, as well as neutrino fluxes
from distant Galactic sources. Let us briefly discuss these
backgrounds separately.

The collision of energetic cosmic rays with nuclei in the upper
atmosphere produces showers of hadrons, mostly pions, that in turn
produce in each decay two muon neutrinos and one electron neutrino.
Neutrinos arrive from all directions at the detector site
after propagating a distance ranging between $\sim 10$ and
12800\usk km while undergoing flavour oscillations. The electron
and muon neutrino fluxes have been carefully computed,
assuming massless neutrinos, by Battistoni et al. with the Monte Carlo
simulation package FLUKA~\cite{Battistoni:2001url}. The theoretical 
uncertainty of the neutrino flux is estimated to be better than
a 20\usk\% over the energy range of interest for this 
paper~\cite{Battistoni:2002ew}.

The effects of neutrino oscillations on the flavour composition of
these fluxes can easily be included using the following
expression for the conversion probability of muon neutrinos
into tau neutrinos:
\begin{equation}
P(\nu_{\mu}\rightarrow\nu_{\tau})=\sin^22\theta_{\text{atm}}\sin^2\left( 1.27\,\frac{\Delta m_{\text{atm}}^2 [\unit{eV}^2] L[\unit{km}]}{E[\unit{GeV}]}\right).
\end{equation}
In this expression, $ E $ is the neutrino energy and
$ L $ is their propagation
length after being produced in the atmosphere, which is given by
\begin{equation}
L=\sqrt{(R_{\oplus}\cos\theta)^2+2R_{\oplus}h+h^2}-R_{\oplus}\cos\theta\,,
\end{equation}
with $ R_{\oplus}=6371\usk\unit{km} $ being the mean Earth radius and
$ h=15\usk\unit{km} $ the mean altitude at which atmospheric muon neutrinos
are produced. Moreover, the
neutrino parameters relevant for the atmospheric oscillations
are $ \sin^22\theta_{\text{atm}}=1 $,
$ \abs{\Delta m_{\text{atm}}^2}=2.4\times 10^{-3}\usk\unit{eV}^2 $~\cite{Schwetz:2008er}. 

In addition to the flux of tau neutrinos originating from the
conversion of muon neutrinos, there exists an intrinsic
contribution from the decay of charmed particles produced in
the atmosphere, coming from all directions, which has a size
about $ 10^6 $ times smaller than
the flux of electron and muon neutrinos from pion decay.
This intrinsic contribution has been computed by Pasquali and
 Reno~\cite{Pasquali:1998xf} and can be parametrised as
\begin{equation}
\log_{10}\left[ E^3\,\frac{dJ_{\nu_{\tau}}}{dE}\left/ \left( \frac{\unit{GeV}^2}{\unit{cm^2\usk s\usk sr}}\right) \right. \right] =-A+Bx-Cx^2-Dx^3,
\end{equation}
where $ x=\log_{10}\left( E[\unit{GeV}]\right)  $, $ A=6.69 $, $ B=1.05 $, $ C=0.150 $ and $ D=-0.00820 $.
The next-to-leading order QCD calculation also shown in their paper gives lower fluxes for
energies below several TeV and is therefore less conservative.

Analogous to the production of neutrinos in the Earth's atmosphere,
neutrinos are produced in the solar corona by cosmic-ray
collisions. This neutrino flux has been studied by
Ingelman and Thunman in~\cite{Ingelman:1996mj},
who found that the flux of electron and muon neutrinos
intergrated over the solar disk can be described by
the following parametrisation:
 \begin{equation}
\frac{d\phi_{x}}{dE} = N_0\,
\frac{\left( E[\unit{GeV}]\right)^{-\gamma-1}}{
1+A \left( E[\unit{GeV}] \right)}\;  (\unit{GeV\usk cm^2\usk s})^{-1},
\end{equation}
which is valid for $ 10^2\usk\unit{GeV} \leq E \leq 10^6\usk\unit{GeV} $.
The numerical values of the coefficients $N_0$, $A$ and $\gamma$
can be found in Tab.~\ref{coronatab} for
$x=\nu_e+\bar{\nu}_e,~\nu_{\mu}+\bar{\nu}_{\mu}$.
\begin{table}
\centering
\begin{tabular}{lccc}
\hline
Flavour & $ N_0 $ & $ \gamma $ & $ A $ \\
\hline
$ \nu_e+\bar{\nu}_e $ & $ 7.4\times 10^{-6} $ & $ 2.03 $ & $ 8.5\times 10^{-6} $  \\
$ \nu_{\mu}+\bar{\nu}_{\mu} $ & $ 1.3\times 10^{-5} $ & $ 1.98 $ & $ 8.5\times 10^{-6} $  \\
\hline
\end{tabular}
\caption{Values for the parametrisation of the corona electron and muon neutrino flux.}
\label{coronatab}
\end{table}
The electron and muon neutrinos and antineutrinos produced
in the solar corona oscillate during their propagation to the Earth.
In view of the long distance travelled, the conversion
and survival probabilities can be averaged, and the fluxes at the Earth
in the different flavours can be straightforwardly calculated
from Eq.~(\ref{oscprob}).

Lastly,
the fluxes of tau neutrinos that originate from Galactic sources are
discussed by Athar, Lee and Lin in~\cite{Athar:2004um}.
For the tau neutrino flux from the Galactic plane in the presence
of neutrino oscillations they find the parametrisation
\begin{equation}
\frac{dJ_{\nu_\tau}}{dE} = 9\times 10^{-6}\usk
(\unit{GeV\usk cm^2\usk s\usk sr})^{-1}\,
\left( E[\unit{GeV}]\right) ^{-2.64},
\end{equation}
which is valid in the energy range
$ 1\usk\unit{GeV}\leq E\leq 10^3\usk\unit{GeV} $.

\section{Detection Prospects}
\label{Detection}

\subsection{Comparison with the Signal}

\begin{figure}
\centering
\includegraphics[scale=1.2]{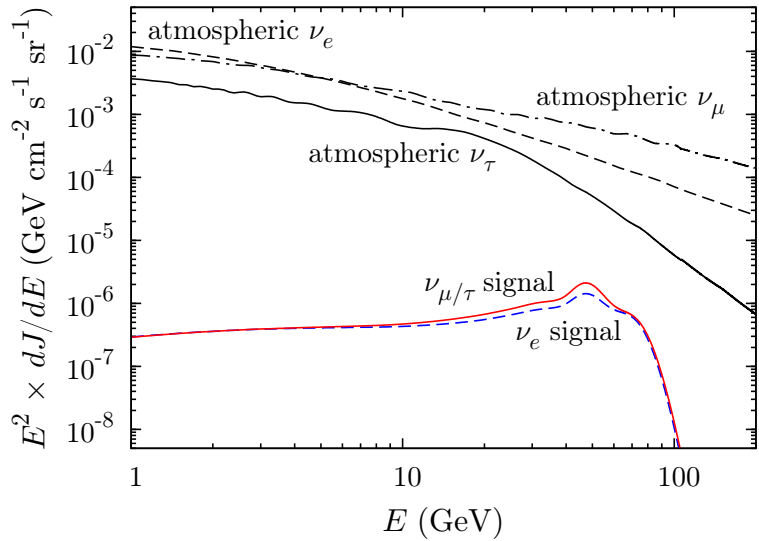}
\caption{Full-sky neutrino fluxes expected at the Super-K site for the
same gravitino mass and lifetime as in Fig.~\ref{eghalo}.
The energy resolution is taken to be 10\usk\%, which is
insufficient to resolve the different peaks in the spectrum.}
\label{nutotal}
\end{figure}

The full-sky signal for the neutrinos from gravitino decay is shown in Fig.~\ref{nutotal} together
with the results for the atmospheric background from FLUKA.
The signal lies several orders of magnitude below the expected atmospheric
background for all flavours. Therefore, we find that the interpretation
of the EGRET and HEAT anomalies in terms of gravitino decay is
compatible with neutrino flux measurements, as it does not lead to
an overproduction of neutrinos.

Going beyond this consistency check, we examine in the following
the possibility of detecting this exotic contribution in neutrino experiments.
Due to the low signal-to-background ratio, the signal cannot be detected
directly. It will therefore be necessary to
find strategies for effectively reducing the background in order to have
any chance of detecting the signal. As is apparent from Fig.~\ref{nutotal},
the tau neutrino channel appears to be the most promising of the
three flavours, since it has the lowest background.
In general, the neutrino spectrum from gravitino
decay has some very specific features that could allow to
distinguish it from the featureless backgrounds, but the question
is whether neutrino detectors will be able to reach sufficient
sensitivity to resolve these features.

\subsection{Electron and Muon Neutrinos}

For the electron and muon neutrinos, which are
more easily detected in neutrino observatories, the signal-to-background ratio is very low
($\sim 10^{-3-4} $) for lifetimes
that are not already excluded by gamma rays~\cite{bbci07,it07}
or antimatter detection~\cite{it08,imm08}.
As can be seen from Fig.~\ref{nutotal}, even the peak of the
spectrum is three orders of magnitude below the background.
This makes distinguishing an exotic signal from the background
extremely difficult.

Unfortunately, we could not find a suitable strategy to sufficiently reduce
this background, e.g. by exploiting directionality. In general the
atmospheric neutrino baseline is too short for all muon
neutrinos to oscillate into another flavour at energies
of order 50--100\usk GeV.
It therefore seems hopeless to try to detect the signal without
having prior knowledge of the position of the peak in the gravitino
decay spectrum.
In case information on the line is available, e.g. from the detection
of a monochromatic gamma-ray line by the Fermi Gamma-Ray Space Telescope
(FGST, formerly named GLAST)~\cite{fgst}, then one could perhaps envisage
strategies to disentangle the signal from the background. However, that
would probably require a much better knowledge of the atmospheric
neutrino flux at the relevant energies and a better energy resolution
than is presently available.

\subsection{Tau Neutrinos}

\begin{figure}
\centering
\includegraphics[scale=1.2]{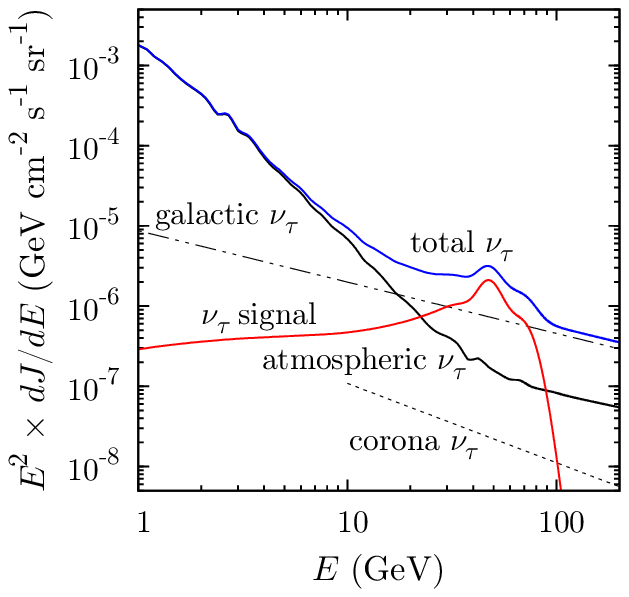}
\includegraphics[scale=1.2]{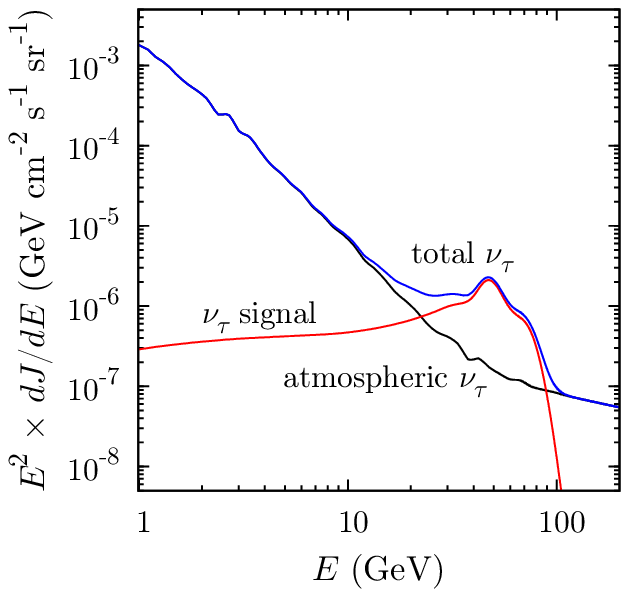}
\caption{Down-going tau neutrino fluxes expected at the Super-K site with (left)
and without (right) the contribution from the Galactic disk and the solar
corona for the same gravitino mass and lifetime as in Fig.~\ref{eghalo}.}
\label{nutaudown}
\end{figure}

For the tau neutrinos, the signal-to-background ratio
is more promising, since it
lies above $\sim 10^{-2}$ at the peak energy.
Moreover, most of the background of tau neutrinos from atmospheric
oscillations and also the other two subdominant sources of tau
neutrinos can be effectively reduced by exploiting directionality.
In fact, the solar corona neutrinos and the Galactic neutrinos
mostly come either from the direction of the Sun or from the Galactic
plane, which could be excluded from the search to reduce the backgrounds.
Furthermore, in our energy
range the tau neutrinos arising from oscillations of the original
muon atmospheric neutrinos are mostly generated for oscillation lengths
of the order of the Earth diameter.
This means that we expect a very low tau neutrino
background if we only consider the flux arriving at the detector
from above the horizon.
In this way, the background of tau neutrinos can be reduced by several
orders of magnitude.
In Fig.~\ref{nutaudown}, we show the fluxes for down-going
tau neutrinos at the Super-Kamiokande site. We see that in this
case, the signal can exceed the simulated background from FLUKA, even without
cutting away the Sun or the Galactic plane.

\subsection{Observability in Super-Kamiokande}

In Cherenkov telescopes, tau neutrinos can only be observed via
charged current (CC) interactions and are very difficult to
disentangle from the other flavours, since the Cherenkov signal
is not sufficiently distinctive to allow identification on an
event-by-event basis.

\begin{figure}
\centering
\includegraphics[scale=1.2]{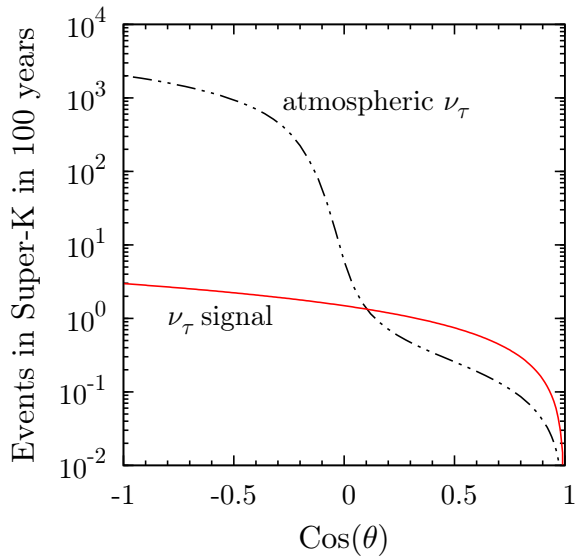}
\caption{Events in 100 years of observation in Super-K due to CC interactions
of $ \nu_{\tau} $ and $ \bar{\nu}_{\tau} $ from the atmosphere (dash-dotted)
and from gravitino decay (solid), integrating the signal from the zenith
direction to the angle $\theta$. The gravitino mass and lifetime are
chosen as in Fig.~\ref{eghalo}.}
\label{CCevents}
\end{figure}

The Super-Kamiokande collaboration has developed a statistical method
to discriminate tau neutrinos from the background of other
flavours~\cite{Abe:2006fu}. Using two different strategies, namely a
likelihood analysis and a neural network, they find an efficiency
of $ 43.1\usk\% $ and $ 39.0\usk\% $, respectively, to identify tau
neutrinos correctly. However, they still misidentify
$ 3.8\usk\% $ and $ 3.4\usk\% $, respectively, of the electron and
muon background neutrinos as tau neutrinos.
Due to the large number of electron and muon neutrino events, the
sample of tau neutrinos is dominated by misidentified neutrinos.
The true tau neutrino events can therefore only be extracted on a statistical
basis using Monte Carlo methods. In the end, the
data is found to be consistent with the atmospheric tau neutrino
flux and neutrino oscillations: The full-sky atmospheric tau neutrino signal 
results in fact in an expected 78 events in the Super-K I period and 43 
events in the Super-K II period~\cite{Abe:2006fu, Kato:2007re}.

However, this analysis does not exploit the information
about the spectral shape of the signal apart from setting a threshold
for $\tau $ lepton production, i.e. 
$ E_{\nu_{\tau}}>m_{\tau}+m_{\tau}^2/2m_n\simeq 3.5\usk\unit{GeV} $,
so this kind of data analysis could certainly be
improved to search for a signal with a peak above the continuum, as
in our case.

Despite the experimental difficulties, it is worthwhile to examine
the theoretically expected signal in the tau channel.
Fig.~\ref{CCevents} shows the expected number of tau
neutrino and antineutrino events per century of observation at Super-Kamiokande
within a zenith angle integrated from $ \cos\theta $ to 1.
If only down-going neutrinos are selected, the signal from
gravitino decays lies above the atmospheric background for higher
energies. However, the fluxes are extremely low and result in
only a few events per century, making it practically impossible
to discriminate them from the other flavours using statistical methods.

One detector specifically optimised for measuring tau neutrinos
above 17\usk GeV event-by-event is OPERA~\cite{opera}, which is already active
in Gran Sasso and will measure tau neutrino appearance in a muon
neutrino beam produced at CERN. Unfortunately, the detector's
effective mass is more than a factor 10 smaller than that of Super-K and
thus, even neglecting the issue of directionality, would be
able to observe only one event from gravitino decays in more than 1000 years.

We therefore conclude that present detectors are unable to
detect the signal, either because they do not have a sufficiently good
efficiency for identifying tau neutrinos, or because they are too
small for the low intensity of our signal, or both.

\section{Future Detectors}
\label{Future}

\begin{figure}
\centering
\includegraphics[scale=1.2]{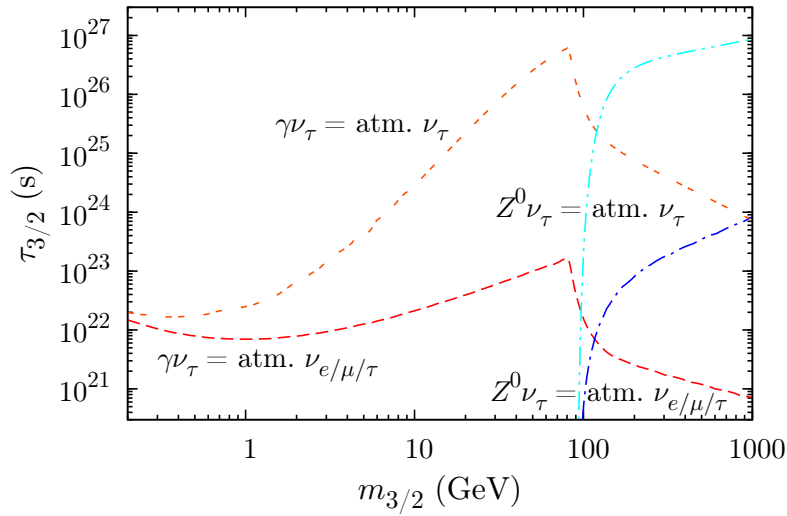}
\caption{Region of the gravitino lifetime where the line signals
from the two-body decays into $\gamma\nu_\tau $ and $ Z^0\nu_\tau $ overcome
the atmospheric background either for all neutrino flavours or for down-going tau
neutrinos only.
We see that considering the tau flavour and exploiting directionality can improve the detectability
by three orders of magnitude for large gravitino masses.
We use here the branching ratios shown in Fig.~\ref{BR-fig}
for the two lines. Note that in the intermediate region the continuum
spectrum from $W^{\pm}$ fragmentation may be used to close the gap,
but we do not consider this possibility here.}
\label{tauvsmass}
\end{figure}

In this section, we will briefly discuss the prospect of detection
for the future. In order to consider the general parameter space
for gravitino decay, we show in Fig.~\ref{tauvsmass} the
region of gravitino lifetime and mass where the
neutrino signal in the $Z^0$ and gamma peaks is equal to the simulated
atmospheric background. For the muon/electron neutrino flux, this
result is similar to that presented in~\cite{PalomaresRuiz:2007ry},
except that we are weighting the channels with the gravitino branching
ratios and that we have only one neutrino produced in the line instead of two. 
Real limits from neutrino experiments have been considered in the past in~\cite{neutrino-old}.

We are taking here as in~\cite{PalomaresRuiz:2007ry} a nominal
energy resolution of 0.3 in $ \log_{10}(E[\unit{GeV}]) $ around the peak
position and we single out the value of the lifetime for which
the peak is equal to the background. Note that since the signal
is proportional to $ 1/\tau_{3/2} $, requiring the peak to be
larger than the background by a specific factor only rescales
the curves by the inverse of this factor.

We clearly see again that the tau neutrino channel in the down-going direction
allows to constrain the gravitino lifetime a few orders of magnitude better
than the muon or electron neutrino fluxes.
On the other hand, similar plots for the gamma-ray channel~\cite{bbci07}
are even more sensitive and give bounds at the order of $10^{27}$\usk s
for gravitino masses below the $W^\pm$ and $Z^0$ thresholds.
For masses above 200\usk GeV, the tau neutrino channel starts to compete
in sensitivity with the photon channel, if we neglect for the moment the
difficulties connected with measuring such a low flux and identifying
the neutrino flavour.
At even higher energies the background flux decreases quickly and
therefore the signal-to-background ratio improves, but the signal rate
then also decreases (as $1/m_{3/2} $), making detection more difficult.

\subsection{Hyper-Kamiokande}

The prospects for Hyper-Kamiokande can be easily obtained by considering
that its mass is planned to be a factor of 10 (for the 0.5 megaton project)
to 20 (for a 1 megaton case) larger than Super-Kamiokande.
Assuming that the rest of the detector performances
are unchanged, we expect to find approximately 20--40 events
from our signal per century from the upper hemisphere.\footnote{
This number could be larger if the fiducial volume of Hyper-Kamiokande
is larger than $\sim 1/2$ the total volume as it is in Super-Kamiokande.}

This number of events might still be too small to allow
for statistical analysis.
However, we expect most of the events to appear within the
peak region or near the threshold and therefore,  an appropriate
energy binning, especially optimised after a signal has already been 
detected in gamma rays, could allow to collect a significant number of
events above the background in a specific energy bin on a shorter
timescale. Still, it is clear that a sufficiently good energy resolution is
a key requirement for singling out the line events, and it remains
uncertain how and if the tau statistical discrimination analysis can
be applied to a sample of such few events.

\subsection{IceCube and km$^3$ Detectors}

Detectors of km$^3$ dimensions have in principle sufficient size
to collect enough events to detect the signal within a
reasonable time span. Even considering
that IceCube is actually looking downwards and
not at the upper hemisphere, from the horizontal direction and the
proton cross-section we estimate ${\cal O}(100) $ events per year
for the completed experiment. Of course, the effective area depends
on the neutrino energy: taking the effective area given
in~\cite{Collaboration:2007rk} for the opposite direction and
assuming most of the signal is above 100\usk GeV, we have instead
${\cal O}(10) $ events per year.
In general, it would be desirable to lower the energy threshold
to reach below 100\usk GeV in order to cover the energy range
favoured by the EGRET and HEAT anomalies.
The combination of IceCube with AMANDA already allows to lower the
threshold to 30\usk GeV. Additionally, plans are being considered for
adding another,
denser subdetector at a deeper location to improve the sensitivity
to dark matter annihilations~\cite{Collaboration:2007rk}.
Such a configuration could probably also be useful for investigating
the present scenario and, more generally, other decaying dark matter candidates.

However, in the case of Cherenkov detectors, the discrimination of tau
neutrinos from other neutrino flavours is generally difficult,
and for IceCube strategies for tau flavour identification have
been proposed only for neutrinos well above TeV energies~\cite{Cowen:2007ny}.
It could therefore be more favourable to improve the energy resolution
and exploit the muon neutrino final state instead.

\section{Conclusions}
\label{Conclusions}

We have examined the neutrino spectrum from the decay of unstable gravitino
dark matter in a scenario with bilinear $R$-parity violation.
It has been pointed out in the recent literature that the decay of gravitino dark matter
particles with a lifetime of $\sim 10^{26}$\usk s
and a mass of $\sim 150 \usk \unit{GeV}$ into massive gauge bosons
may account for the
anomalies observed in the diffuse extragalactic gamma-ray spectrum
as measured by EGRET as
well as in the positron fraction as measured by HEAT.\footnote{The existence
of a positron excess seems to be supported
by preliminary results from PAMELA~\cite{pamela08}.}
Motivated by this observation, we have computed the neutrino
spectrum for the same choice of parameters as a consistency check 
of this scenario. We find that this spectrum is compatible with results from
neutrino experiments. 

We have also examined the detectability of this exotic component of the 
neutrino flux to find an independent way to test this scenario.
While the signal in the neutrino spectrum with two or more distinct peaks,
resulting from two-body gravitino decays into gauge/Higgs boson and neutrino,
is very characteristic, it will be challenging to detect these features in 
neutrino experiments.
On one side, present neutrino detectors do not achieve a sufficiently high 
energy resolution to resolve the subdominant peaks, and on the other side,
the event rate is expected to be so small that the background of atmospheric 
neutrinos overwhelms the signal in all flavours.
The most promising signal-to-background ratio is found in the
tau neutrino flavour, especially when analysing only the flux from
the upper hemisphere since there the atmospheric tau neutrino
flux is vastly reduced. However, tau neutrinos are difficult to identify 
in Cherenkov detectors and probably only an event-by-event identification 
procedure could allow the signal to be seen with such extremely limited statistics.
At present, therefore, it is not possible to detect this contribution due 
to technological limitations.

The ideal detector for testing the present scenario would be one of 
megaton mass 
with the ability to identify and measure tau neutrinos event by event. 
Should such a detector ever become available, it could be worthwhile to look 
for this component of the neutrino flux by employing strategies for 
background reduction such as the ones discussed here, especially 
if the anomalous signatures in the positron fraction and the diffuse 
extragalactic gamma-ray spectrum are confirmed by PAMELA and 
FGST, respectively.
The detection of a signal in neutrinos compatible with signals in 
the other indirect detection channels would in fact bring significant 
support to the scenario of decaying dark matter, possibly consisting of 
gravitinos that are unstable due to bilinear $R$-parity violation.

\section*{Acknowledgements}

We would like to thank Wilfried Buchm\"uller, Marco Cirelli,
Concha Gonz\'alez-Garc\'ia,
Ricard Tom\`as and Mark Vagins for useful discussions.

LC would like to thank NORDITA and the organisers of the
NORDITA program on "TeV scale physics and dark matter" for
hospitality and support during part of this work.
LC also acknowledges the support of the "Impuls- und Vernetzungsfond"
of the Helmholtz Association under the contract number VH-NG-006 and
of the European Network of Theoretical Astroparticle Physics ILIAS/N6
under contract number RII3-CT-2004-506222. The work of AI and DT
was partially supported by the DFG cluster of excellence Origin and 
Structure of the Universe.

\end{document}